\documentstyle[psfig,conf_iap]{article}
\def\puncspace{\ifmmode\,\else{\ifcat.\C{\if.\C\else\if,\C\else\if?\C\else%
\if:\C\else\if;\C\else\if-\C\else\if)\C\else\if/\C\else\if]\C\else\if'\C%
\else\space\fi\fi\fi\fi\fi\fi\fi\fi\fi\fi}%
\else\if\empty\C\else\if\space\C\else\space\fi\fi\fi}\fi}
\def\SP{\let\\=\empty\futurelet\C\puncspace}

\def\h1{$h^{-1}$\SP}

\def\etal{{\it et al.\/}\ }

\def\eg{{\it e.g.\/}\rm,\ }
\def\lsim{~\rlap{$<$}{\lower 1.0ex\hbox{$\sim$}}}
\def\gsim{~\rlap{$>$}{\lower 1.0ex\hbox{$\sim$}}}
\def\void#1{{}}

\begin{document}
\heading{%
%
ESO Imaging Survey: Finding Targets for VLT
%
} 
\par\medskip\noindent
\author{%
Luiz Nicolaci da Costa
}
\address{%
European Southern Observatory, Karl-Schwarzschild-Str 2, D-85748
Garching b. M\"unchen, Germany
}

%

\begin{abstract}
Data from the wide-angle, moderately deep ESO Imaging Survey have been
used to produce target lists for the first year of the VLT. About 250
candidate clusters of galaxies have been identified from the $I-$band
images covering $\sim$ 17 square degrees. In addition, using the
multicolor data available over an area of 1.3 square degrees over 300
potentially interesting point-sources have been selected. The
color-selected targets include low-mass stars/brown dwarfs,
white-dwarfs and quasars. Images, object catalogs and derived target
lists are available from the world-wide web (http://www.eso.org/eis).
\end{abstract}

\section{Introduction}

The ESO Imaging Survey (EIS) was conceived as a public survey to provide
the ESO community with suitable data for producing target lists for the
first year of VLT operation~\cite{Renzini}. EIS has been carried out as a
concerted effort of ESO and its community to optimize the one-year
period available between the re-commissioning of the NTT in July 1997 and
the deadline for proposals for the first period of scientific operation
of the VLT. To maximize the return a new framework was established by
ESO and the Observing Program Committee whereby a Working Group was
created to define the goals and to oversee the execution of a public
imaging survey. A Visitor Program was also created to allow tapping on
the expertise of different European groups to develop the tools required
for the efficient translation of raw images into potentially useful
target lists. As a by-product the project has also been used to
establish the infra-structure (pipeline processing, archive,
object-oriented database) required to cope with the large increase in
the data flow expected from new CCD-mosaic cameras under development
for La Silla and Paranal, fully dedicated for wide-field imaging.

The first phase of EIS (EIS-wide) consisted of a moderately deep
($I\lsim 23.5$) wide-angle survey, covering four patches of sky spread
over the right ascension range $22^h<\alpha<9^h$, thus providing targets
nearly year-round.  The observations of EIS-wide have already been
completed and all the data in the form of photometrically and
astrometrically calibrated pixel maps, object and derived catalogs are
publicly available. These data can be retrieved from the web via an user
interface built in collaboration with the ESO Science Archive group as a
prototype for future distribution of public data. Table 1 summarizes the
position of the EIS patches and the area covered in each of the
passbands considered.

\begin{center}
{\bf Table 1.} EIS-Wide Sky Coverage
\end{center}
\begin{center}

\begin {tabular}{lcccccc}
\hline
Patch   & $\alpha$ & $\delta$& B  & V   & I  \\
\hline
A       & 22:42:54  &   -39:57:32 & -   & 1.2  & 3.2 \\
B       &  00:49:25 &   -29:35:34 & 1.7  & 1.7 & 1.8  \\
C       &  05:38:24  &  -23:51:00  &  -   &  -   & 6.0  \\
D       &   09:51:36 &   -21:00:00 &-   &  -   & 6.0 \\
\hline
        &    -       &      -      & 1.7  & 2.9  & 17  \\
\hline
\end{tabular}
\end{center}

A full description of the EIS pipeline and the quality of the data for
each patch can be found in papers that accompanied each data
release~\cite{Nonino}~\cite{Prandoni}~\cite{Benoist}. The computed
star counts are, in general, consistent with model predictions, while
the galaxy counts obtained for each patch are internally consistent
and in good agreement with those of other authors. Moreover, we find
that the $I-$band observations are sufficiently deep to search for
distant clusters. Internal and external comparisons of the galaxy
angular two-point correlation function are also in good agreement,
indicating that the derived galaxy catalogs are uniform.

At the time of this writing, we have already started the second phase of
EIS (EIS-deep) which consists of a deep, multicolor survey in five
optical and three infrared passbands covering 75 arcmin$^2$ of the
HST/Hubble Deep Field South (HDFS), including the WFPC2, STIS and NICMOS
fields, and a region of 100~arcmin$^2$ selected for deep X-ray
observations with AXAF. The data will be used to find U- and B-dropouts
and to produce galaxy samples with photometric redshifts from which 
galaxies in the redshift range $ 1 < z < 2$ can be drawn for follow-up
spectroscopic observations in the infrared.

\section{Cluster Candidates}

Cluster candidates were identified using a cluster finding pipeline
implemented at the back-end of the EIS pipeline~\cite{Olsena}. The
search algorithm is based on the matched-filter
technique~\cite{Postman} and it was chosen to facilitate the
comparison with the results of one of the few systematic searches for
optically-selected distant clusters~\cite{Postman}. It should be
emphasized that the primary goal of the EIS team has been to prepare a
list of cluster {\it candidates} for follow-up observations and not to
produce a well-defined sample for statistical analysis. Instead, our
main concern has been to minimize the number of false detections,
thereby increasing the yield in future follow-up work. For this
purpose the analysis has been restricted to the most uniform surveyed
areas and parameters were chosen conservatively, using an extensive
set of simulations. We point out that the derived catalog is not
unique, given the various underlying assumptions of the method and our
particular choice of parameters. However, since the data are public,
other groups may produce their own catalogs using different methods
(\eg Lobo, this conference). A comparison of various catalogs will be
instructive in evaluating the strengths and weaknesses of different
algorithms.

\begin{figure} \centerline{\vbox{
\psfig{figure=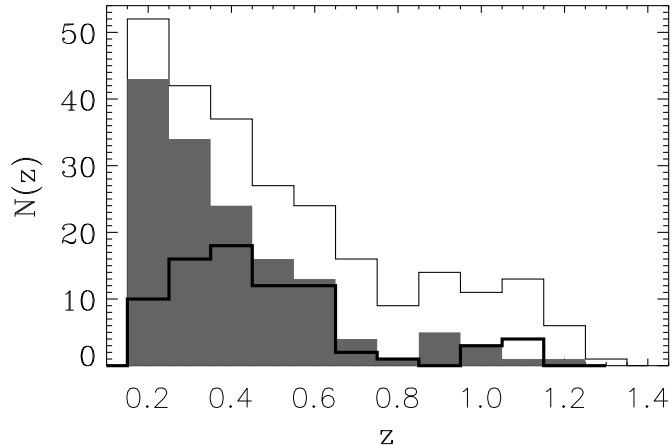,height=7.cm} }} \caption[]{The redshift
distribution of the total sample of EIS clusters covering an area of
$\sim 14.4$ square degrees. The shaded histogram corresponds to
detections with high significance. The thick line shows the estimated
redshift distribution of cluster candidates in the PDCS, covering 5.1
square degrees.} \label{fig:zall} \end{figure}

The total sample of EIS $I-$band cluster candidates consists of 252
objects in the redshift range $0.2 \leq z \leq 1.3$~\cite{Scodeggio}.
The redshift distribution of the sample is shown in
figure~\ref{fig:zall}. The median redshift of the distribution is $z\sim
0.4$. Note that the EIS redshift distribution differs somewhat from that
observed by PDCS, also shown in the figure. The number of EIS candidates
decreases monotonically with redshift up to $z \sim 0.6$ with an
extended tail beyond, in contrast to the PDCS which shows a relatively
flat distribution peaking at $z \sim 0.4$.

Another way of further testing the reality of the detections is to use
data in different passbands. Using the $V$-band data available for
$\sim$ 2.7 square degrees, clusters were identified and
cross-correlated to the $I$ detections. The results can be summarized
as follows~\cite{Olsenb}: 1) About 90\% of the cluster
candidates with $z \le 0.5$ and about 25\% with $z>0.5$, primarily
rich clusters, are confirmed using the $V$ candidates; 2) Candidates
at low-redshift show the red envelope in the C-M relation expected for
ellipticals. The CM relation serves as an independent confirmation of
the candidate clusters and an independent redshift estimate, by and
large consistent with the estimates from the matched filter method.

\section {Color-Selected Targets}

Preliminary lists of other potentially interesting targets were also
extracted from the multicolor data obtained for a 1.7 square degree
region near the South Galactic Pole.  The region was observed in $B, V$
and $I$ and offers a unique combination of area and depth.  These lists
contain a total of 358 objects ($I \lsim 21.5$) over 1.27 square
degrees, after eliminating regions observed under less than ideal 
conditions. Among the color selected targets are candidate very low mass
stars/brown dwarfs (62), white-dwarfs (32), and quasars
(264)~\cite{Zaggia}. These objects are natural candidates for follow-up
spectroscopic observations and illustrate the usefulness of the EIS data
for a broad range of science. The selected objects can be found in the
web and can be examined by displaying side-by-side image postage stamps
in the three passbands. Improvements in the sample selection are
certainly possible and we encourage interested groups to produce their
own samples.  From the present data the derived samples  typically
include  50 to 100 candidates each. However, much larger samples will be
available from the Pilot Survey to be carried out with the new
wide-field camera on the 2.2 m telescope at La Silla.

\section{Summary}

One year after the first observations all the data for EIS-wide are
available to the community.  In addition, a basic pipeline for the
processing of images is available which allows for the processing of
individual frames, the coaddition of overlapping images, the
extraction of objects and the preparation of color catalogs.  The
pipeline is currently being generalized to handle data from from
different detectors and CCD mosaics, an essential step for its use
with the wide-field imager WFI@2.2m to be commissioned later this
year.

The EIS candidate cluster catalog consists of about 250 candidates in
the redshift range $0.2 < z < 1.3$, distributed over a wide range of
right ascension, with candidates available year-round. This sample is
currently the largest available in the southern hemisphere and
confirmation work will start soon. EIS has also produced over 300
color-selected targets for different scientific goals. More
importantly, these samples can grow in time if similar public surveys
are carried out on the 2.2m telescope, with an efficiency at least 6
times higher than the original EIS at the NTT.  It should be point out
that a Pilot Survey on the 2.2m, which aims at completing EIS-wide,
has already been scheduled for the beginning of 1999.

\acknowledgements{I would like to thank all the
EIS team members for the magnificent work carried out over the
past year. Without their dedication and commitment it would have been
impossible to meet the strict deadline under which this survey was
conducted.  Special thanks to Alvio Renzini for his unconditional
support and enthusiasm. Finally, we would like to thank ESO's Director
General Riccardo Giacconi for making this effort possible and for his
and P. Rosati's choice of the AXAF field.}


%
\vfill
\end{document}